\documentclass{article}
\usepackage{arxiv}
\usepackage[utf8]{inputenc} % allow utf-8 input
\DeclareUnicodeCharacter{0300}{`{}}
\DeclareUnicodeCharacter{0302}{`{}}
\usepackage[T1]{fontenc}    % use 8-bit T1 fonts
\usepackage{hyperref}       % hyperlinks
\usepackage{url}            % simple URL typesetting
\usepackage{booktabs}       % professional-quality tables
\usepackage{amsfonts}       % blackboard math symbols
\usepackage{nicefrac}       % compact symbols for 1/2, etc.
\usepackage{microtype}      % microtypography
\usepackage{lipsum}
\usepackage{graphicx}
\usepackage{caption}
\usepackage{float}
\usepackage{dsfont}
\usepackage{xcolor}
\usepackage{mathtools}

\usepackage{amsmath}

\usepackage{amssymb}
\usepackage{authblk}
\usepackage{colortbl}
\usepackage{multicol}
\usepackage{multirow}
\usepackage{biblatex}
\bibliography{references}

\usepackage{fontawesome}
\usepackage{hyperref}
\usepackage{fancyhdr}
\usepackage{tabularray}
\usepackage{svg}
\usepackage{latexsym}
\usepackage{framed,multirow}
\usepackage{subcaption}  % For subfigures
\usepackage[bottom]{footmisc}

\newcommand*\samethanks[1][\value{footnote}]{\footnotemark[Equal contribution]}

\title{4D-VQ-GAN: Synthesising Medical Scans at Any Time Point for Personalised Disease Progression Modelling of Idiopathic Pulmonary Fibrosis}

\author[1,2]{An Zhao*}
\author[1,6]{Moucheng Xu*}
\author[1,2]{Ahmed H. Shahin}
\author[3]{Wim Wuyts}
\author[4]{Mark G. Jones} 
\author[1,5]{Joseph Jacob}
\author[1,2]{Daniel C. Alexander}

\affil[1]{Hawkes Institute, University College London}
\affil[2]{Department of Computer Science, University College London}
\affil[3]{Department of Respiratory Medicine, University Hospitals Leuven}
\affil[4]{NIHR Southampton Biomedical Research Centre and Clinical and Experimental Sciences, University of Southampton}
\affil[5]{UCL Respiratory, University College London}
\affil[6]{Department of Med. Physics \& Biomedical Engineering, University College London}
\affil[*]{Equal contribution}
\affil[\faPhone]{Contact: an.zhao.19@alumni.ucl.ac.uk, moucheng.xu.18@alumni.ucl.ac.uk}

\begin{document}

\maketitle

\begin{abstract}

Understanding the progression trajectories of diseases is crucial for early diagnosis and effective treatment planning. This is especially vital for life-threatening conditions such as Idiopathic Pulmonary Fibrosis (IPF), a chronic, progressive lung disease with a prognosis comparable to many cancers. Computed tomography (CT) imaging has been established as a reliable diagnostic tool for IPF. Accurately predicting future CT scans of early-stage IPF patients can aid in developing better treatment strategies, thereby improving survival outcomes. In this paper, we propose 4D Vector Quantised Generative Adversarial Networks (4D-VQ-GAN), a model capable of generating realistic CT volumes of IPF patients at any time point. The model is trained using a two-stage approach. In the first stage, a 3D-VQ-GAN is trained to reconstruct CT volumes. In the second stage, a Neural Ordinary Differential Equation (ODE) based temporal model is trained to capture the temporal dynamics of the quantised embeddings generated by the encoder in the first stage. We evaluate different configurations of our model for generating longitudinal CT scans and compare the results against ground truth data, both quantitatively and qualitatively. For validation, we conduct survival analysis using imaging biomarkers derived from generated CT scans and achieve a C-index comparable to that of biomarkers derived from the real CT scans. The survival analysis results demonstrate the potential clinical utility inherent to generated longitudinal CT scans, showing that they can reliably predict survival outcomes.

\end{abstract}

\begin{keywords} {4D image synthesis \and VQ-GAN \and neural ODEs \and spatial-temporal disease progression modelling \and CT \and IPF}
\end{keywords}

\section{Introduction}
Disease progression modelling aims at discovering disease evolution trajectories and is crucial for understanding diseases' biological mechanisms. While significant progress has been made in building disease progression models using longitudinal, low-dimensional data (e.g., differential equation models \cite{villemagne2013amyloid,jack2013brain,oxtoby2014learning} and self-modelling regression methods \cite{jedynak2012computational,donohue2014estimating}), a significant challenge remains: developing such models for high-dimensional volumetric imaging data. This task involves using a limited number of observations to generate imaging data for any given time point to mimic the disease progression trajectory. Compared to low-dimensional data, volumetric scans offer a wealth of information, but at the cost of increased computational complexity. This area remains relatively unexplored. However, promising initial attempts exist, leveraging generative models to simulate longitudinal magnetic resonance imaging (MRI) scans specifically in the context of Alzheimer's disease (AD) progression \cite{couronne2021longitudinal,sauty2022progression, kim2021longitudinal,marti2023mc,fan2022tr,puglisi2024enhancing,ravi2022degenerative,yoon2023sadm}.

Beyond neurodegenerative diseases, disease progression models have also been investigated in progressive lung conditions such as idiopathic pulmonary fibrosis (IPF). However, there remains a need for models capable of generating longitudinal CT scans for IPF using limited observational data. Existing methods for generating longitudinal MRI scans in AD are not directly applicable to IPF for several reasons. First, IPF is much rarer with shorter patient lifespans than AD resulting in less imaging being acquired. Furthermore, radiation side effects from CT scans discourage repeated scans, resulting in a scarcity of longitudinal CT data and hindering model development. Second, lung CT scans contain vastly more fine textured-structures like vessels, airways, and interstitial tissue compared to the smooth, homogenous brain seen in MRI. Generating these intricate lung structures synthetically is more challenging than replicating brain tissue.

Several problems need to be tackled to address this task. Modelling disease progression directly in the image space of volumetric CT scans is challenging due to the complexity of lung structures. A more effective approach is to use autoencoder-based frameworks to map the images to latent space for progression modelling, then project them back to the original image space \cite{sauty2022progression,kim2021longitudinal,marti2023mc,fan2022tr,puglisi2024enhancing}. Another challenge lies in determining how to model the dynamics of disease progression. Some previous disease progression models \cite{sauty2022progression,kim2021longitudinal} rely on explicit assumptions about evolution trajectories (e.g. linear progression). This oversimplification fails to capture the complexity observed in real-world disease progression. Other methods \cite{marti2023mc,fan2022tr} use recurrent neural networks to capture the temporal information. These methods model dynamics in discrete steps, limiting their ability to capture the continuous nature of disease progression. Puglisi et al.\cite{puglisi2024enhancing} use the latent diffusion model and incorporate prior knowledge to model the disease progression. However, it also struggles to ensure continuous temporal changes. To overcome the limitations of these methods, Neural Ordinary Differential Equations (ODEs) emerge as a promising alternative \cite{chen2018neural}. Neural ODEs provide a powerful framework for modelling continuous disease progression by learning the underlying dynamics of the disease state. They achieve this by parameterizing the derivative of the hidden state with neural networks, allowing them to capture complex, non-linear progression with greater flexibility.

In this paper, we introduce the 4D Vector Quantized Generative Adversarial Network (4D-VQ-GAN). Given 3D imaging data at irregular time points, our method can generate synthetic 3D images at any desired time point, effectively modelling a continuous disease progression trajectory for each individual. We demonstrate the effectiveness of our approach in generating CT volumes for IPF patients. Additionally, we found that biomarkers derived from the generated CT volumes exhibit a strong clinical correlation with survival outcome, highlighting the potential of our method for personalized treatment planning.

\section{Methods}
\label{sec:methods}
\begin{figure}[!t]
\centering
\includegraphics[width=0.8\textwidth]{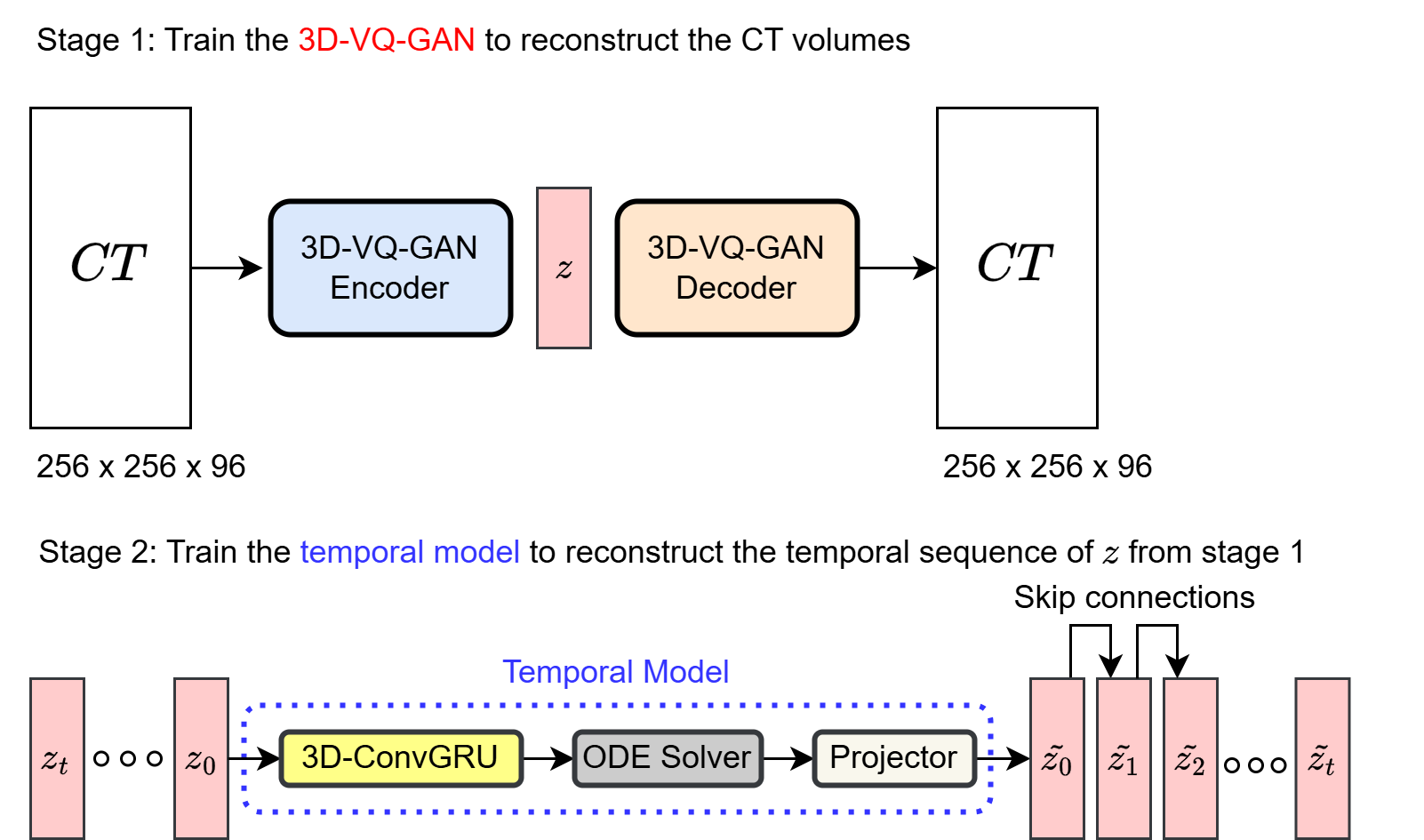}
\caption{The overview of our two-stage training strategies. The first stage trains an encoder-decoder-based 3D-VQ-GAN to reconstruct the CT volumes. The second stage takes the latent embeddings ($z_t, ..., z_0$) from the first stage, and trains a temporal model to reconstruct them. The temporal model consists of a 3D-ConvGRU, that compresses the temporal latent embeddings to match the dimensionality of the input of the ODE Solver to ease the computational burden. The projector, a light-weight 3D convolutional module reconstructs the temporal latent embeddings from the outputs of the ODE Solver. Those reconstructed latent embeddings are then fed into a frozen 3D-VQ-GAN Decoder from stage 1 for longitudinal CT reconstruction.}  
\label{fig:4dvqgan}
\end{figure}

Our 4D-VQ-GAN is a self-supervised generative model trained on temporal CT volumes. The model consists of two key components: a 3D-VQ-GAN and a temporal model, trained in a two-stage process. During inference, given two CT scans of the same patient at different time points, our model can generate new CT volumes at any desired time point, effectively capturing disease progression. In this section, we first introduce these components, as shown in \ref{fig:4dvqgan}, and present our proposed survival analysis method to evaluate the effectiveness of the generated CT scans in predicting patients' future outcomes.

\begin{figure}[!t]
\centering
\includegraphics[width=0.8\textwidth]{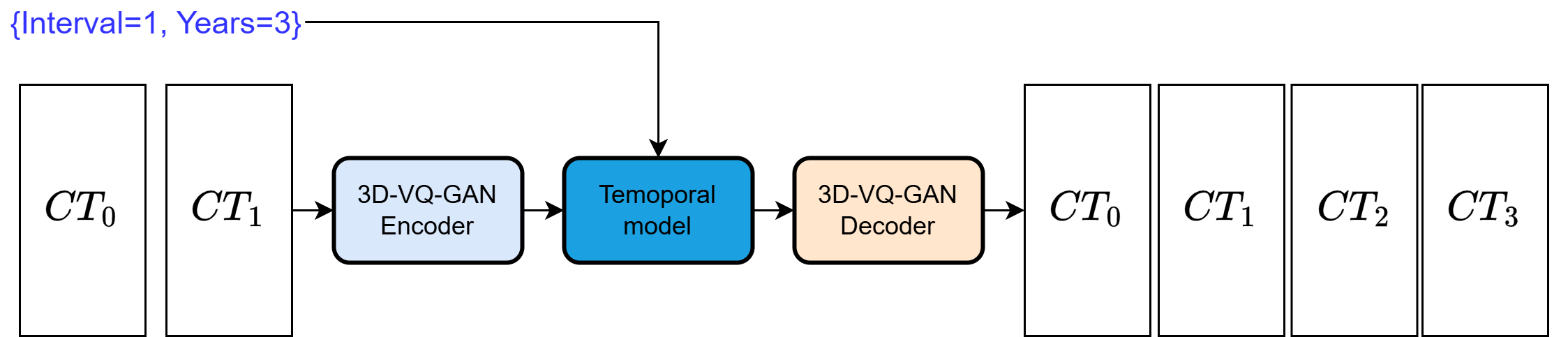}
\caption{The inference of the trained model. Given two scans, the model can generate more scans up to the specified year at a specific interval.}  
\label{fig:inference}
\end{figure}

\textbf{3D-VQ-GAN} As shown in \ref{fig:4dvqgan}, the first stage of training involves a 3D-VQ-GAN \cite{ge2022long} to reconstruct CT volumes for each case at every time point in the training set. Unlike the original 2D-VQ-GAN \cite{esser2021taming}, our 3D-VQ-GAN employs 3D convolutional layers, enabling it to capture spatial structures more effectively in volumetric imaging data. Following the VQ-VAE framework \cite{van2017neural}, 3D-VQ-GAN compresses high-dimensional volumetric imaging data into a discrete set of latent codes, constrained by a predefined codebook size. Since the model is trained on imaging data, its codebook learns to represent meaningful imaging patterns, with each code acting as a compact and discrete representation of local anatomical structures or texture features. We adapted the original 3D-VQ-GAN loss functions \cite{ge2022long} for training. 
Further details can be found in Appendix \ref{appendix:3dvqgan}.

\textbf{Temporal Model} In the second stage of the training, we train a temporal model that can reconstruct the temporal trajectories of the latent embeddings ($z$) of the imaging data from different time-points, as demonstrated in \ref{fig:4dvqgan}. We realise that generating the future or past latent embeddings from a few observed latent embeddings naturally formulates as an ordinary differential equation. We therefore utilise a neural ODE solver \cite{chen2018neural} to predict the unknown embeddings at new time-points. We found that it is more beneficial to adapt a 3D-ConvGRU \cite{ballas2015delving} as the encoder of the neural ODE solver on our data, with much better computational efficiency. The outputs of the neural ODE are then fed into a light-weight projector, two 3D convolutional layers, to reconstruct the latent embeddings $z$ at inquired time points, with a specified time interval. In practice, we also discovered that adding skip-connections for the generated embeddings gives better results. This is because, by adding the skip-connections between consecutive latent embeddings, the generation task became easier, as the model only needs to learn the differences between the latent embeddings at every two adjacent time-points. We train the temporal model in a self-supervised manner by using a L2 loss between the input embeddings and the reconstructed embeddings. 
Please refer to Appendix \ref{appendix:temporal} for more details.

\textbf{Inference} The inference follows the process shown in \ref{fig:inference}. The trained model needs two CT scans. The user can also input two hyper-parameters for the neural ODE, namely the interval time and the total time duration. For example, as shown in \ref{fig:inference}, given two initial scans at time point 0 and 1, the interval as 1 year and the total time duration as 3 years, the model will reconstruct the scans at time point 0 and 1, and start to extrapolate future scans at time point 2 and 3. 

\textbf{Survival analysis and biomarker discovery} To evaluate the potential clinical utility of the proposed method, we perform survival analysis based on scans generated by the trained model. As discussed before, each CT scan can be represented as a set of codebook indices in the latent space, with the most meaningful features of the scan embedded within these indices. Each code index represents a distinct imaging pattern in the CT. The frequency of each code index reflects the prevalence of the corresponding imaging pattern. We then use these normalized frequencies of code indices as candidate prognostic biomarkers. The survival analysis is conducted in three stages. In the first stage, we identify the top five embedded imaging biomarkers from the above candidate biomarkers. Our biomarker selection process goes as follows. For each candidate biomarker, we input it along with relevant covariates (e.g., sex, age, smoking status) from the training dataset into the Cox proportional hazards model \cite{cox1972regression}. The biomarkers are then ranked by their p-values, and the top five are selected for further analysis. In the second stage, we input these five selected biomarkers, along with the covariates, into the Cox model to compute the final C-index for both real and generated future scans in the test dataset. This step validates whether significant prognostic information is preserved in the generated future scans. These biomarkers are derived from a single CT scan and are thus considered cross-sectional biomarkers. In the third stage, we obtain the longitudinal biomarkers from both real and generated CT scans by calculating the changes of the top five biomarkers over a one-year period. These biomarkers, along with the covariates, are input into the Cox model to assess their prognostic value in the test dataset. The third stage highlights the potential utility of tracking changes in these biomarkers over time.

\begin{figure}[!t]
\centering
\includegraphics[width=0.9\textwidth]{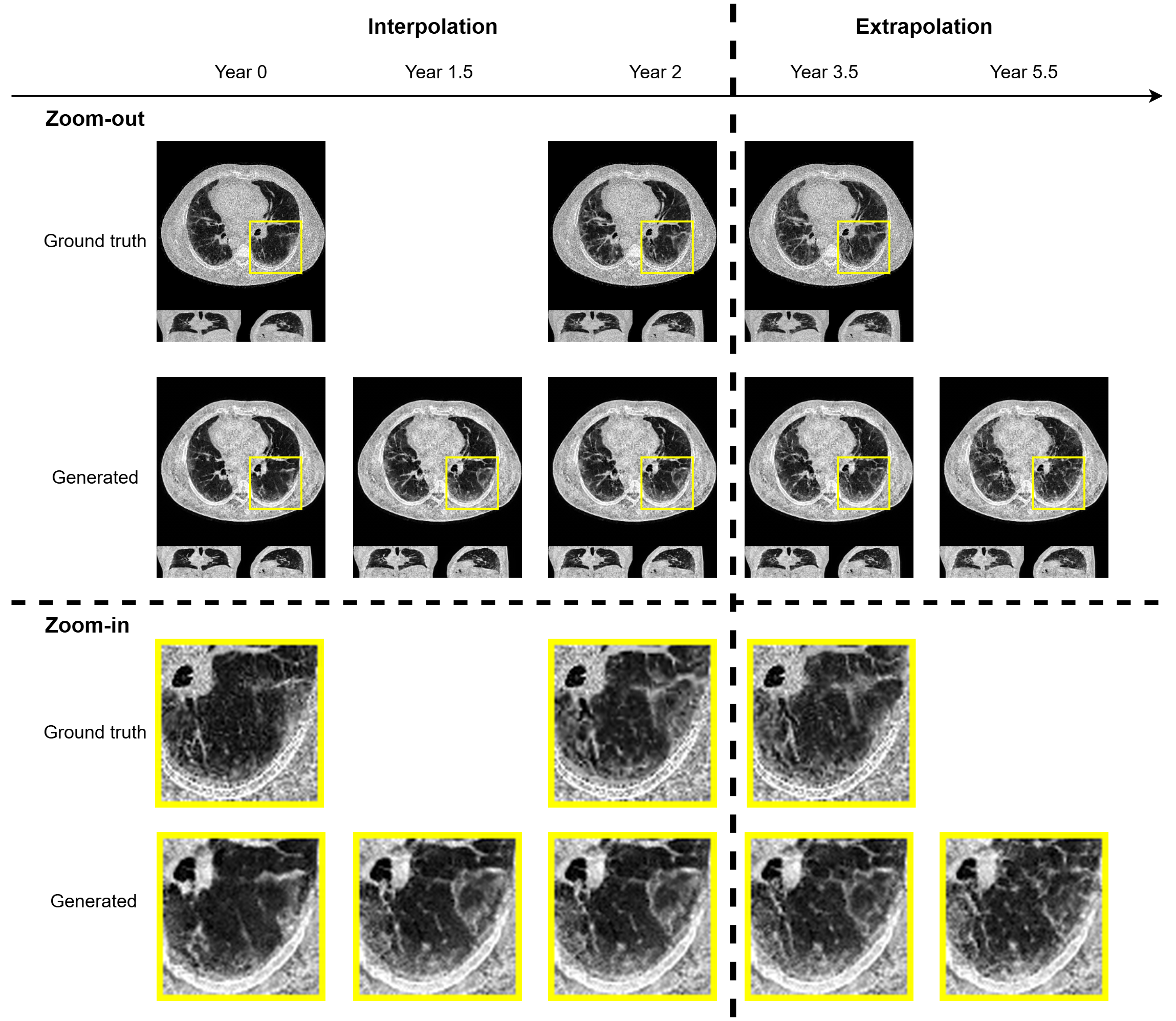}
\caption{Three real CT scans of an IPF patient are shown in the upper panel, representing axial, coronal, and sagittal sections. Using two scans from year 0 and year 2, the trained model can generate CT scans at any arbitrary time points. The below panel shows the generated CT images at five different time points, with three corresponding to the real scans. A zoomed region of the left lower lobe (yellow box) in the real and generated CT scans show comparable amounts of architectural distortion, patterned ground glass opacification and reticulation, all hallmarks of lung fibrosis.} 
\label{fig:example}
\end{figure}

\section{Experiments}
\textbf{Datasets} Our data comes from a longitudinal dataset comprising 681 volumetric CT scans from 219 IPF patients, obtained from University Hospitals Leuven, Belgium, a single centre in Leuven. We randomly divided the dataset into 80\% for training (552 CT scans from 175 patients) and 20\% for validation (129 CT scans from 44 patients). The CT scans were acquired with varying slice thicknesses, ranging from 0.75 mm to 1 mm, and in-plane resolution varied with pixel spacing ranging from 0.38 mm to 0.98 mm. Additionally, we use an external cross-sectional dataset of 98 IPF patients from the University Hospital Southampton NHS Foundation Trust, UK, to evaluate the generalizability of the 3D-VQ-GAN trained in Stage 1 and more details can be found in Appendix \ref{appendix:reconstruction}. 

\textbf{Preprocessing} We focus on modelling changes within the lung areas. We segment the lung regions using a pre-trained U-Net~\cite{ronneberger2015u,hofmanninger2020automatic} for all CT scans of IPF patients and visually inspect the lung masks. Subsequently, we register the longitudinal lung scans to remove extraneous artifacts caused by lung motion or incorrect body positioning. Our lung scan registration method is a faster version (implemented in \cite{hansen2021graphregnet}) of the CorrField method \cite{heinrich2015estimating}. Visualizations of the segmentation and registration outcomes can be found in Appendix \ref{appendix:Pre-processing}.

\textbf{Training} Our models are trained on an NVIDIA A100 80GB GPU. The first training stage used a batch size of 1, with an accumulated batch size of 6, and lasted 20,000 steps, equivalent to 10 days. The second training stage also used a batch size of 1 and lasted 15 hours. The hyperparameters of the training can be found in Appendix \ref{appendix:training_details}.

\textbf{Evaluation Metrics} To evaluate the image quality of the reconstructed CT scans in the first stage of training, we use Mean Squared Error (MSE). For assessing the image quality of the generated temporal CT scans in the second stage, we use Peak Signal-to-Noise Ratio (PSNR) and Structural Similarity Index (SSIM). For survival analysis, we use the Concordance Index (C-index), a metric that measures the predictive accuracy of a model by evaluating the agreement between predicted risk scores and actual survival outcomes.

\section{Results}
In this section, we present the results of our model on the following tasks: 1) interpolation, for imputing missing CT scans between the two input scans; 2) extrapolation, for predicting the future CT scans beyond the time span of the two given input scans; 3) survival outcome prediction, for evaluating of the clinical utility of the proposed method.

\begin{table}[!ht]
\centering
\caption{Ablation Study on the architecture of the temporal model.}
\label{tab5_2}
\begin{tabular}{lcccccc}
\toprule
Model Variant & \multicolumn{3}{c}{Interpolation} & \multicolumn{3}{c}{Extrapolation} \\
\cmidrule(lr){2-4} \cmidrule(lr){5-7}
 & MSE$\downarrow$ & SSIM$\uparrow$ & PSNR$\uparrow$ & MSE$\downarrow$ & SSIM$\uparrow$ & PSNR$\uparrow$ \\
\midrule
\scriptsize ODE encoder (ConvGRU) & 0.018 & 0.500 & 17.913 & 0.023 & 0.472 & 16.974 \\
\scriptsize ODE encoder (ConvGRU)+embeddings skip connections (Ours) & 0.020 & 0.469 & 17.387 & 0.019 & 0.489 & 17.816 \\
\scriptsize ODE encoder (ODE-ConvGRU)+embeddings skip connections & 0.018 & 0.508 & 17.900 & 0.026 & 0.440 & 16.913 \\
\scriptsize ODE encoder (ConvGRU)+embeddings skip connections+masked out inputs & 0.035 & 0.343 & 15.366 & 0.036 & 0.340 & 15.240 \\
\bottomrule
\end{tabular}
\end{table}

\textbf{Interpolation and extrapolation} 
In this experiment, we evaluate the performance of the proposed method in reconstructing latent dynamics from sparse, irregularly sampled data. The model is trained to reconstruct the given input sequence of 3D imaging data and is then tested on both interpolation and extrapolation tasks. Figure \ref{fig:example} illustrates example qualitative results. The patient's year 0 and year 2 scans were used for inference, with the year 3.5 scan serving as ground truth for the extrapolation experiment. Inference was performed with a 0.5-year interval, extending up to 5.5 years from the baseline (year 0). As shown in the zoomed-in section, both real and generated CT scans exhibit comparable architectural distortion, ground glass opacification, and reticulation—key features of lung fibrosis. The quantitative results are shown in \ref{tab5_2}.

\textbf{Ablation study} 
The proposed method uses Neural ODE to predict the difference embeddings for each time point relative to the previous one, incorporating skip connections. In the ablation experiments (\ref{tab5_2}), we remove the skip connections and allow the model to directly predict the latent embeddings of all time points using Neural ODE. However, this modification resulted in significantly poorer extrapolation performance. We also replaced ConvGRU with ODE-ConvGRU \cite{park2021vid} as the ODE solver encoder. While ODE-ConvGRU is designed for irregularly sampled longitudinal data, it leads to notably worse extrapolation performance compared to the model using standard ConvGRU. Additionally, we experiment with randomly masking time points (excluding the baseline) and task the model with reconstructing the unseen time points. This approach also results in significantly worse extrapolation performance. Overall, the proposed model demonstrates balanced performance in both interpolation and extrapolation tasks.

\textbf{Survival outcome prediction} 
Following the method in Section \ref{sec:methods}, the top five most significant cross-sectional imaging biomarkers are selected for survival analysis. In the test dataset, we extrapolate the third CT scan using the first two available CT scans. For the cross-sectional imaging biomarker,
the C-index using the generated third scans is 0.886. In comparison, using biomarkers derived from real CT scans for survival prediction yields a better C-index of 0.929. Next, we compute the longitudinal biomarker by evaluating the change in these top five significant biomarkers over the course of one year, both for real and generated CT scans. Inputting these longitudinal biomarkers along with covariates into the Cox model results in C-indices of 1.0 for real CT scans and 0.959 for generated CT scans. However, these results may be overestimated due to the limited sample size. The C-index of survival analysis using generated scans is comparable to that of real scans, further highlighting the strong clinical potential of our method.

\begin{table}
\caption{Survival analysis results.}\label{tab_survival_analysis}
\centering
\resizebox{0.7\textwidth}{!}
{%
\begin{tblr}{
  cell{1}{1} = {r=2}{},
  cell{1}{2} = {c=2}{c},
  cell{3}{2} = {r},
  cell{3}{3} = {r},
  cell{4}{2} = {r},
  cell{4}{3} = {r},
  vlines,
  hline{1,3-5} = {-}{},
  hline{2} = {2-3}{},
}
Biomarker type                 & C-Index                          &                                      \\
                          & Real CT scans &  Generated CT scans (ours) \\
Cross-sectional biomarker & 0.929                            & 0.886                                \\
Longitudinal biomarkers   & 1                                & 0.959                                
\end{tblr}}
\end{table}

\section{Discussion and Conclusion}
This paper represents a pioneering effort in building a deep-learning-based disease progression model to generate 4D imaging data continuously, mimicking disease progression trajectories of IPF patients from sparse discrete observations. By integrating 3D-VQ-GAN and Neural ODE, the proposed method avoids the strong linear assumptions commonly used in previous disease progression models for 3D imaging data. This enables the modelling of more complex trajectories and the generation of synthetic 3D imaging data, which is crucial for capturing the spatial and temporal heterogeneity of diseases like IPF. Being able to reliably predict evolution and disease progression in IPF is extremely challenging. Today the main guide to disease progression is monitoring lung function test trajectories which are associated with measurement variability and may be unreliable in the presence of other lung pathology such as emphysema. Accordingly, using imaging as shown, may help transform clinical management decisions such as initiating medication or expediting referral for lung transplantation.

Beyond disease progression analysis, the model has the potential for applications like synthetic data generation, data augmentation, and missing data imputation. However, the proposed method exhibits certain limitations. Firstly, the current approach utilizes a deterministic Neural ODE in the latent space, which assumes a shared disease dynamic across all patients. This might not be ideal for heterogeneous diseases with diverse subtypes and distinct progression patterns. Secondly, a crucial limitation of the current validation process is its reliance on a single dataset for evaluating the model. Future work should address these limitations by incorporating diverse datasets and disease dynamics, to refine the model and pave the way for its potential clinical applications. Our model is designed to be scalable, allowing future work to also explore larger-scale implementations. With additional pre-training and computational resources, the proposed method could generate even more refined results.

\printbibliography
% \clearpage

\appendix
\section{Pre-processing}
\label{appendix:Pre-processing}

The registration process aligns corresponding structures across scans, ensuring the model focuses on disease-related changes. The Learn2Reg challenge, associated with MICCAI 2020 and 2021, provides valuable insights for choosing a suitable registration method for lung CT scans.  This challenge compared various approaches on clinically relevant tasks. In the lung CT task, the goal was to register expiration scans to inspiration scans, with corresponding landmarks provided for accuracy evaluation \cite{hering2022learn2reg}. Notably, CorrField \cite{heinrich2015estimating}, a non-rigid registration method, emerged as the top performer among 15 methods, including deep learning and conventional approaches. CorrField achieved a target registration error (the Euclidean distance between corresponding landmarks in the warped fixed and moving scan) of only 1.75mm \cite{hering2022learn2reg}.

We use the faster version (implemented in \cite{hansen2021graphregnet}) of CorrField method \cite{heinrich2015estimating}, a non-learning-based unsupervised method. CorrField first employs Foerstner operator \cite{forstner1987fast} to extract distinctive keypoints in one 3D volume. Then a dissimilarity distribution over a densely quantized space of displacements is calculated. Finally, a parts-based model is used to infer the smooth motion of connected keypoints and regularize the correspondence field. Specifically, a minimum spanning tree (MST) is generated from the set of sparse keypoints, which enables exact message passing using belief propagation on the graph to regularise the displacement costs. To ensure the disease progression model focuses solely on lung tissue, we define keypoints within lung masks for registration. After aligning the longitudinal CT scans to the baseline CT scan using image registration, we replaced the non-lung region in the registered scans with the corresponding region from the baseline CT scan. This eliminates distractions from surrounding body parts. We leverage the default hyperparameters from the validated implementation (\url{https://grand-challenge.org/algorithms/corrfield/}) for registration. To verify registration quality, all registered scans undergo visual inspection by me to identify and exclude those with significant errors. However, because of the non-clinical background, there is still risk compared with verification conducted by experienced radiologists.

\begin{figure}[H]
\includegraphics[width=0.9\textwidth]{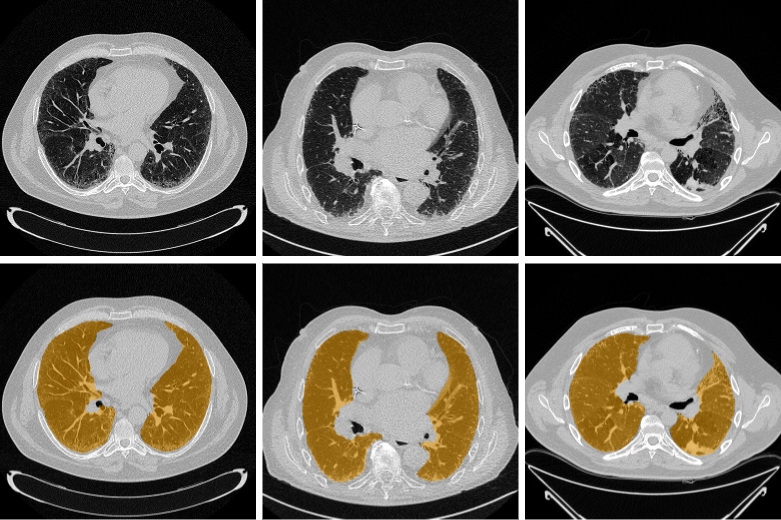}
\caption{Segmentation results for selected cases from Leuven cohort}\label{supp_fig_3_1}
\end{figure}

\begin{figure}[H]
\includegraphics[width=0.9\textwidth]{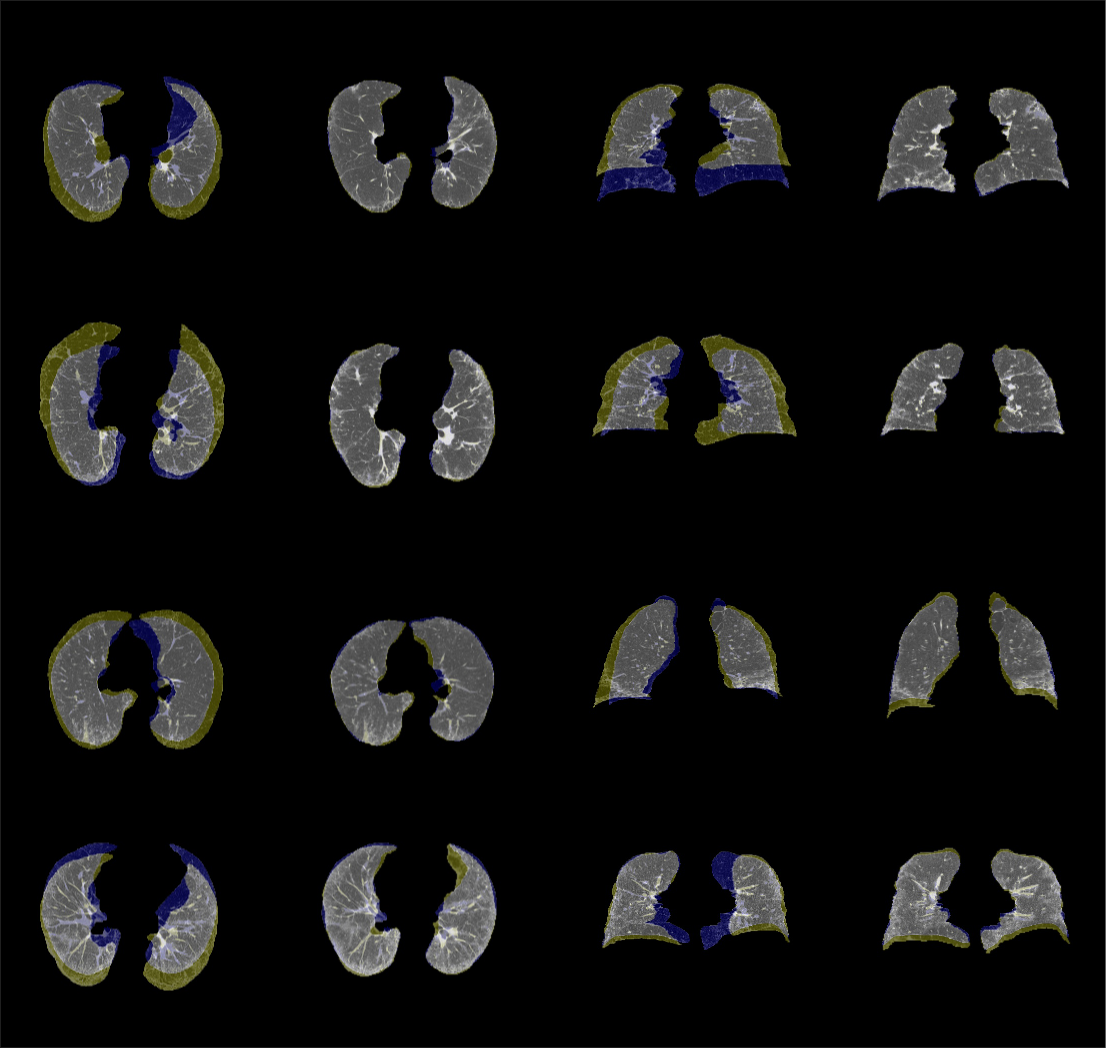}
\caption{Visualization of four registration outcomes with a focus on lung areas for clarity. The left two columns present axial views before and after registration, while the right two columns showcase coronal views. Baseline scans are denoted in blue, whereas follow-up scans are highlighted in yellow. The merging of colours results in grey or white hues, indicating aligned structures due to RGB amalgamation. Notably, follow-up scans are registered to their corresponding baseline scans. The first two rows illustrate cases with successful registration outcomes, while the subsequent two rows demonstrate instances of varying degrees of misalignment.}\label{supp_fig_3_2}
\end{figure}

\section{Additional experiments}
In this section, we began by conducting experiments on the initial stage of the model to show the reconstruction performance of 3D-VQ-GAN, both qualitatively and quantitatively. Additionally, we explore the impact of varying hyperparameters on the reconstruction performance. Subsequently, we assess the proposed two-stage model's effectiveness in disease progression modelling and interpolation tasks, demonstrating its capability to capture the dynamics of disease progression. Finally, we present visualizations of the learned codebook for enhanced interpretability.

\begin{table}[H]
\centering
\caption{Reconstruction performance of 3D-VQ-GAN with different hyperparameters.}
\label{tab5_1}
\resizebox{\textwidth}{!}{%
\begin{tblr}{
  cell{2}{2} = {c},
  cell{2}{3} = {c},
  cell{2}{4} = {c},
  cell{2}{5} = {c},
  cell{2}{6} = {c},
  cell{3}{2} = {c},
  cell{3}{3} = {c},
  cell{3}{4} = {c},
  cell{3}{5} = {c},
  cell{3}{6} = {c},
  cell{4}{2} = {c},
  cell{4}{3} = {c},
  cell{4}{4} = {c},
  cell{4}{5} = {c},
  cell{4}{6} = {c},
  cell{5}{2} = {c},
  cell{5}{3} = {c},
  cell{5}{4} = {c},
  cell{5}{5} = {c},
  cell{5}{6} = {c},
  cell{6}{2} = {c},
  cell{6}{3} = {c},
  cell{6}{4} = {c},
  cell{6}{5} = {c},
  cell{6}{6} = {c},
  hlines,
  vlines,
}
No. & Vocabulary size & Compression rate & Learning rate  & MSE on internal test set & MSE on external test set \\
1   & 1024            & 4                & $3\times10^{-4}$ & $3.76\times10^{-3}$        & $6.84\times10^{-3}$        \\
2   & 4096            & 4                & $3\times10^{-4}$ & $3.54\times10^{-3}$        & $6.64\times10^{-3}$        \\
3   & 256             & 4                & $1\times10^{-4}$ & $4.88\times10^{-3}$        & $8.67\times10^{-3}$        \\
4   & 256             & 4                & $3\times10^{-4}$ & $4.23\times10^{-3}$        & $8.10\times10^{-3}$        \\
5   & 256             & 8                & $3\times10^{-4}$ & $6.17\times10^{-3}$        & $1.15\times10^{-2}$        
\end{tblr}}
\end{table}

\begin{figure}
\includegraphics[width=0.9\textwidth]{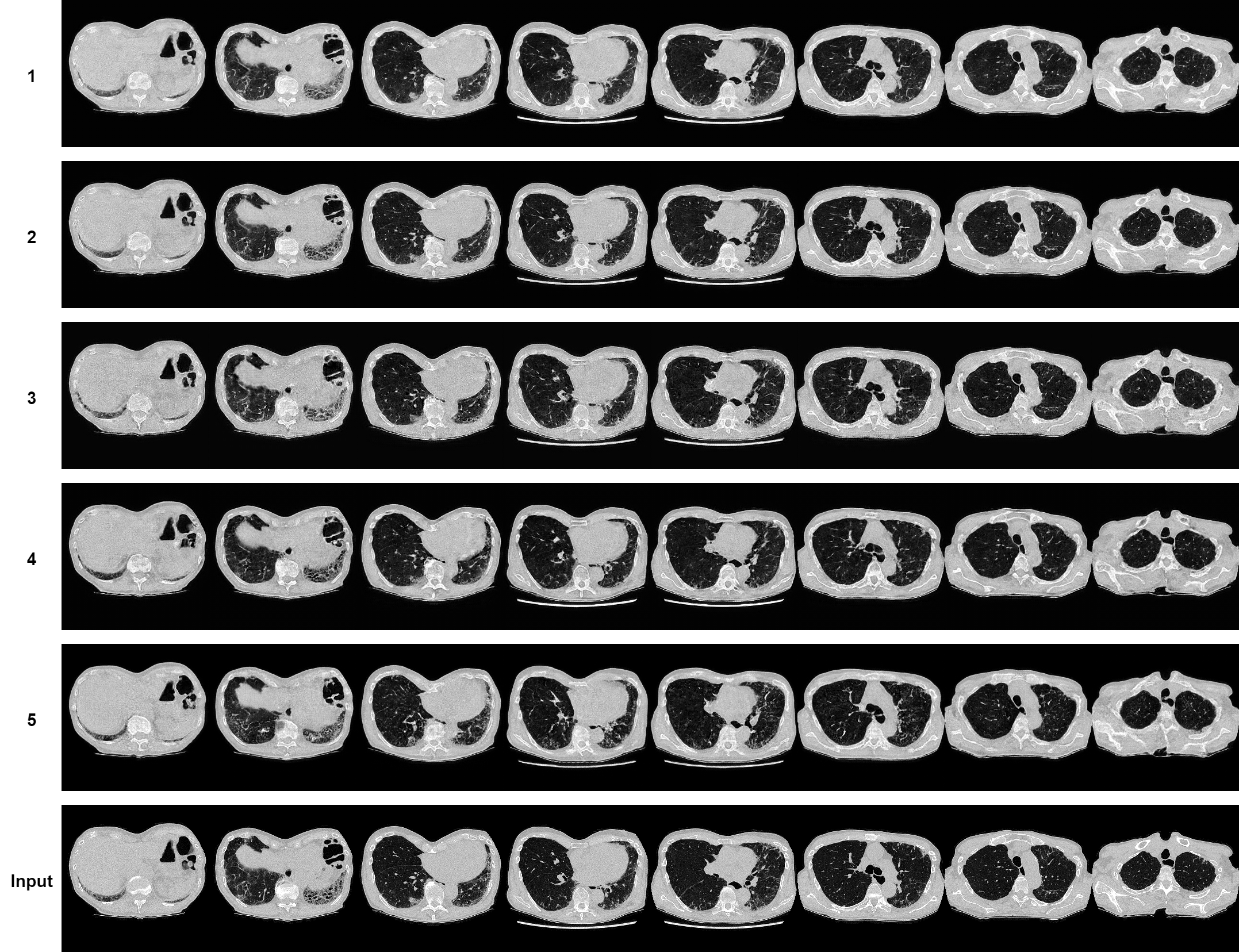}
\caption{Examples of input 3D CT scans and corresponding CT scans reconstructed by 3D-VQ-GAN with varying hyperparameters, as detailed in \ref{tab5_1}} \label{fig5_2}
\end{figure}

\section{Training and implementation  details}
\label{appendix:training_details}
We train the model for two stages, 3D-VQ-GAN and latent disease ODE. All models are trained on a NVIDIA A100 80GB GPU.
\subsection{3D-VQ-GAN}
In line with recommendations from \cite{ge2022long, esser2021taming}, the training of 3D-VQ-GAN begins on all CT scans in the training set using the reconstruction loss. Subsequently, the GAN loss is introduced after 10,000 steps. Hyperparameters are set as follows: $\lambda_{\textrm{perc}}=\lambda_{\textrm{rec}}=4$ and $\lambda_{\textrm{GAN}}=1$. The Adam optimizer \cite{kingma2014adam} is employed with a learning rate of $3\times10^{-4}$ and $\beta_1=0.5, \beta_2=0.9$. Training the 3D-VQ-GAN spans 20,000 steps, and the best model, determined by the smallest training loss after adding the GAN loss, is selected. The batch size is set at 1, and the accumulated batch size is 6. Training the first-stage model takes approximately 10 days.
\subsection{Disease ODE}
After completing the training of 3D-VQ-GAN, we proceed to train the latent disease ODE using the AdamW optimizer \cite{loshchilov2017decoupled}. The training spans 100 epochs, employing a batch size of 1, a learning rate of $2\times10^{-4}$, and $\beta_1=0.5, \beta_2=0.9$. To enhance the model's performance at later time points, we implement a linearly increasing weights strategy, assigning higher loss weights for later time points. The best model, identified by the smallest training loss, is selected. Training the latent disease ODE takes approximately 15 hours.
\subsection{Implementation details}
For the implementation of 3D-VQ-GAN, we adopt a similar network structure as outlined in \cite{ge2022long}. The codebook size is set to $M=256$ with an embedding size of $c=16$. We employ a compression rate $r=4$, calculated as the ratio between $D, H, W$ and $d, h, w$. All input 3D CT scans are resized to $D=96, H=256, W=256$ before being fed into the model. In the latent disease ODE, the neural ODE solver is implemented by stacking three 3D convolution layers. The code is implemented using PyTorch 1.8.

\section{Details of the Methodology}
\subsection{Problem formulation}
We denote the irregularly sampled longitudinal 3D imaging data of a subject as $\mathcal{X_{\mathcal{T}}} = \{\mathbf{X}_{t_0},\mathbf{X}_{t_1},...,\mathbf{X}_{t_j},...,\mathbf{X}_{t_T}\}$, where each $\mathbf{X}_{t_j}\in \mathbb{R}^{D\times H\times W \times C}$ ($D$: depth, $H$: height, $W$: width, $C$: channel) and $\mathcal{T}=\{t_0,t_1,...,t_j,...t_T\}$ ($t_j$: the time of the $j$\textsuperscript{th} observation of the subject). Each subject can have an arbitrary number of observations. In the context of IPF disease progression modelling, this translates to each patient having an arbitrary number of longitudinal volumetric lung CT scans. The corresponding mask for the region of interest is also segmented. Given $\mathcal{X_{\mathcal{T}}}$, the objective of this work is to build a model capable of generating synthetic 3D imaging data at any time point between $t_0$ and $t_T$, illustrating the progression of the disease within a patient over time. In this study, we use 3D volumetric CT scans of IPF patients as an example application.

\begin{figure}[H]
\centering
\includegraphics[width=0.9\textwidth]{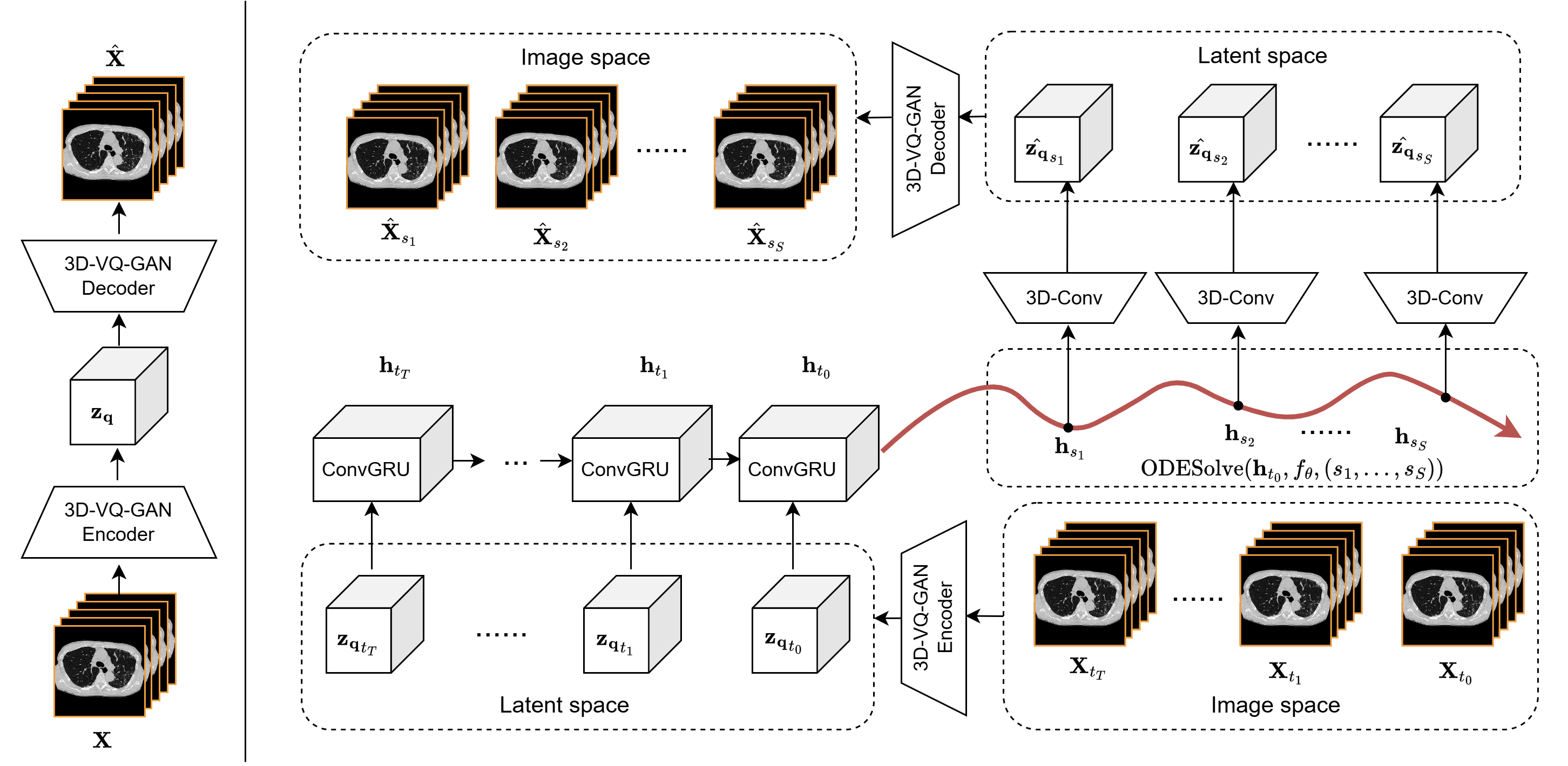}
\caption{The overview of the proposed two-stage model. The left side is the 3D-VQ-GAN for image reconstruction. The right side is the latent disease ODE for modelling disease progression dynamics from longitudinal 3D imaging data.}
\label{fig5_1}
\end{figure}

The proposed method has two stages: In the first stage, a 3D-VQ-GAN is trained to reconstruct CT volumes. In the second stage, a latent (ODE) is trained to model the temporal dynamics from quantised embeddings of longitudinal CT scans generated by the encoder in the first stage, reconstructing continuous trajectories from discrete observations in the latent space (Figure \ref{fig5_1}).

\subsection{3D-VQ-GAN for image reconstruction}
\label{appendix:3dvqgan}

\begin{figure}[H]
\centering
\includegraphics[width=0.9\textwidth]{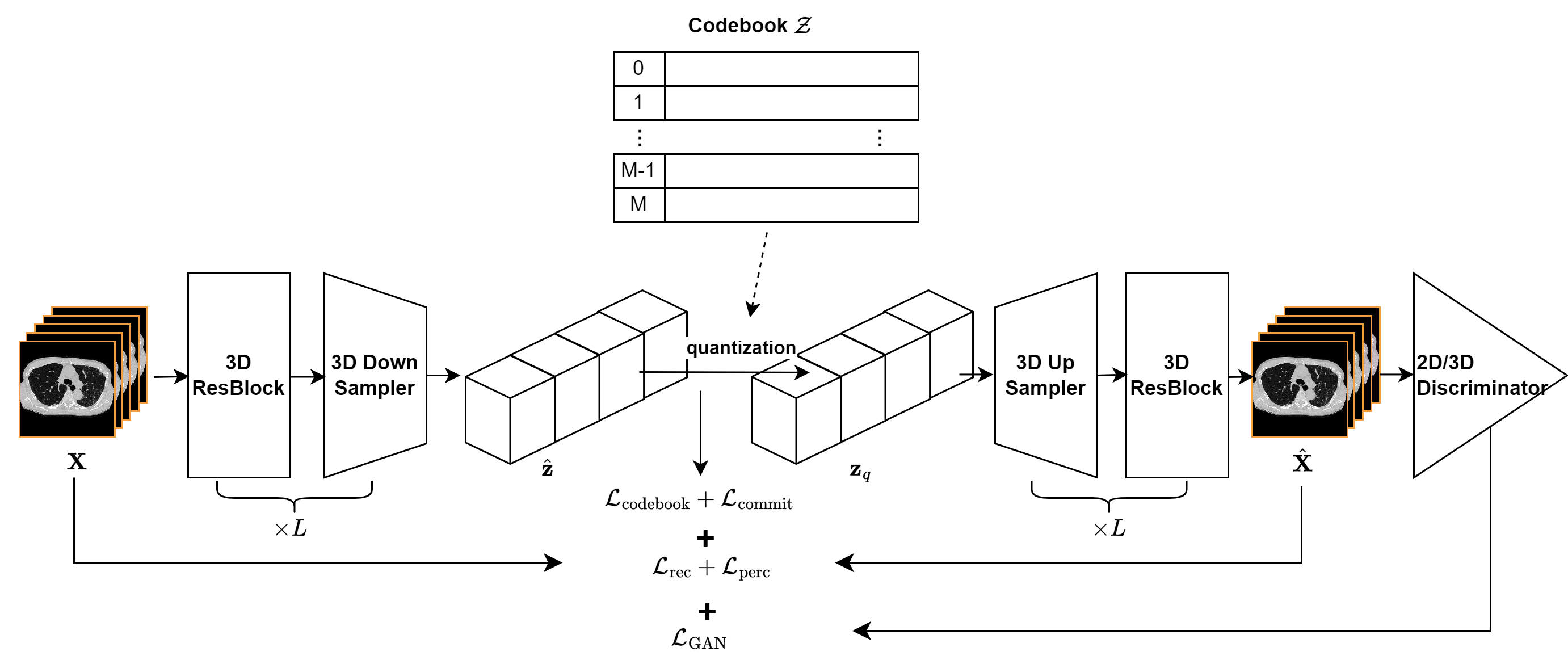}
\caption{The overview of the 3D-VQ-GAN for image reconstruction.}
\label{3D-VQ-GAN}
\end{figure}

To capture the dynamics of disease progression in the latent space, which can significantly decrease computational costs, the model needs to first learn an effective and compact latent representation of the input image. In the first stage, we adopt 3D-VQ-GAN \cite{ge2022long} (Figure \ref{3D-VQ-GAN}), which replaces 2D operations of the original VQ-GAN \cite{esser2021taming} with 3D operations. VQ-GAN is a variant of VQ-VAE. VQ-VAE consists of an encoder $E$ and a decoder $G$ and keeps a discrete codebook of learned representations in latent space. Given an input $\mathbf{X}$, the VQ-VAE tries to represent the input image with embeddings from the codebook in the latent space. More specifically, the encoder $E$ projects $\mathbf{X}$ into the embedding $\hat{\mathbf{z}}=E(\mathbf{X}) \in \mathbb{R}^{d \times h \times w \times c}$ in latent space followed by an element-wise quantization operation $\mathbf{q}(\cdot)$ which approximates $\hat{\mathbf{z}}$ by replacing each spatial code $\hat{\mathbf{z}}_{i,j,k} \in \mathbb{R}^c$ with its nearest neighbour in the trainable codebook $\mathcal{Z}=\{\mathbf{z}_m\}_{m=1}^M\in \mathbb{R}^c$. The discrete latent indices and embeddings after quantization are denoted as $\mathbf{c} \in \mathbb{Z}^{d \times h \times w}$ and $\mathbf{z_q} \in \mathbb{R}^{d\times h\times w\times c}$ respectively. $\mathbf{z_q}$ then goes through the decoder to reconstruct the input $\hat{\mathbf{X}}=G(\mathbf{z_q})$, using: $\mathbf{z_q} = \mathbf{q}(\hat{\mathbf{z}})=\underset{\mathbf{z}_m \in \mathcal{Z}}{\arg\min}||\hat{\mathbf{z}}_{i,j,k}-\mathbf{z}_m||$.

For the non-differentiable quantization operation, VQ-VAE uses a straight-through estimator \cite{bengio2013estimating} which copies gradients from decoder input $\mathbf{z_q}$ to encoder output $\mathbf{\hat{z}}$ \cite{van2017neural}. 

\begin{equation}
    \mathbf{z_q} = \mathbf{q}(\hat{\mathbf{z}})=\underset{\mathbf{z}_m \in \mathcal{Z}}{\arg\min}||\hat{\mathbf{z}}_{i,j,k}-\mathbf{z}_m|| 
\end{equation}

The VQ-VAE loss $\mathcal{L}_{\rm{vqvae}}$ consists of three terms: reconstruction loss $\mathcal{L}_{\textrm{rec}}$, codebook loss $\mathcal{L}_{\textrm{codebook}}$ and commitment loss $\mathcal{L}_{\textrm{commit}}$. $\mathcal{L}_{\textrm{rec}}$ is used for optimizing both encoder and decoder. $\mathcal{L}_{\textrm{codebook}}$ is used only for optimizing the codebook by pushing embeddings in the codebook to be close to the output of the encoder. $\mathcal{L}_{\textrm{commit}}$ is employed to enforce the encoder commits to an embedding in the codebook \cite{van2017neural}.

\begin{equation}
\mathcal{L}_{\rm{vqvae}} = \underbrace{||\mathbf{X}-\hat{\mathbf{X}}||_1}_{\mathcal{L}_{\textrm{rec}}} + \underbrace{||\textrm{sg}[E(\mathbf{X})]-\mathbf{z_q}||_2^2}_{\mathcal{L}_{\textrm{codebook}}}+\underbrace{\beta||\textrm{sg}[\mathbf{z_q}]-E(\mathbf{X})||_2^2}_{\mathcal{L}_{\textrm{commit}}}
\end{equation}

$\rm{sg}[\cdot]$ is the stop-gradient operator here.

In addition to the VQ-VAE loss, VQ-GAN also uses GAN loss $\mathcal{L}_{\textrm{GAN}}$ and perceptual loss $\mathcal{L}_{\textrm{perc}}$ to improve the reconstruction quality as well as increase the compression rate. Similar to 3D-VQ-GAN \cite{ge2022long}, we use two discriminators $D_{\textrm{2d}}$ and $D_{\textrm{3d}}$. $D_{\textrm{2d}}$ is used to distinguish the real slice and the reconstructed slice of 2D plane. $D_{\textrm{3d}}$ is used to distinguish real 3D input $\mathbf{X}$ and reconstruction $\mathbf{\hat{X}}$ to encourage the consistency between slices:

\begin{equation}
\mathcal{L}_{\textrm{GAN}}=\log{D_{\textrm{2d/3d}}(\mathbf{X})+\log(1-D_{\textrm{2d/3d}}(\hat{\mathbf{X}})})
\end{equation}

Perceptual loss \cite{zhang2018unreasonable} measures the distance of the true input and reconstruction in the feature space of a VGG network \cite{simonyan2014very}.
\begin{equation}
\mathcal{L}_{\textrm{perc}}=\sum_l{w_l||\textrm{VGG}^{(l)}(\mathbf{\hat{X}})-\textrm{VGG}^{(l)}(\mathbf{X})||_1}
\end{equation}

$\textrm{VGG}^{(l)}(\cdot)$ extracted the features of $l^{th}$ layer of VGG. $w_l$ is a learned weight for scaling.

The overall loss of 3D-VQ-GAN $\mathcal{L_\textrm{3D-VQ-GAN}}$ would be
\begin{equation}
\begin{aligned}
    &\min_{E,G,\mathcal{Z}}{(\max_{D_{\textrm{2d}},D_{\textrm{3d}}}{(\lambda_{\textrm{GAN}} \mathcal{L}_{\textrm{GAN}}}}))+\\
    &\min_{E,G,\mathcal{Z}}{(\lambda_{\textrm{perc}} \mathcal{L}_{\textrm{perc}}+\lambda_{\textrm{rec}} \mathcal{L}_{\textrm{rec}}+\mathcal{L}_{\textrm{codebook}}+\beta\mathcal{L}_{\textrm{commit}})}
\end{aligned}
\end{equation}

\subsection{Disease ODE: modelling latent disease progression dynamics}
\label{appendix:temporal}
\subsubsection{Overview}
As shown in Figure \ref{fig5_1}, given the trained encoder of 3D-VQ-GAN, longitudinal input 3D imaging data ${\mathbf{X}_{t_0},\mathbf{X}_{t_1},...,\mathbf{X}_{t_T}}$ for each patient can be projected to a series of quantized embeddings ${\mathbf{z_\mathbf{q}}_{t_0},\mathbf{z_\mathbf{q}}_{t_1},...,\mathbf{z_\mathbf{q}}_{t_T}}$. This sequence of embeddings can be considered as samples from the continuous disease trajectory of that subject in the embedding space. To reconstruct the continuous trajectory from discrete observations, we adapt the common latent ODE structure, an encoder-decoder-based latent-variable time series model. In this application, the primary focus lies on capturing changes within specific Regions of Interest (ROIs), i.e. lung area. To isolate and emphasize these areas in the analysis, we apply a masking technique that excludes regions outside of the ROI in the latent embedding series. The overview of the latent disease ODE is shown in Figure \ref{fig5_1}. Firstly, we use convolution-based gated recurrent unit (3D-ConvGRU) neural network \cite{ballas2015delving} as an encoder to embed the input sequence ${\mathbf{z_\mathbf{q}}_{t_0},\mathbf{z_\mathbf{q}}_{t_1},...,\mathbf{z_\mathbf{q}}_{t_T}}$ into a latent initial state $\mathbf{h}_{t_0}$. Then, the continuous latent trajectory can be generated by using an ODE solver given $\mathbf{h}_{t_0}$. Finally, the latent trajectory is projected back to the embedding space of 3D-VQ-GAN to get embeddings ${\hat{\mathbf{z_\mathbf{q}}}_{s_1},\hat{\mathbf{z_\mathbf{q}}}_{s_2},...,\hat{\mathbf{z_\mathbf{q}}}_{s_S}}$at any target timesteps $\mathcal{S}=\{s_1,s_2,...,s_S\}$. Feeding this sequence to the trained decoder $G$ of 3D-VQ-GAN can reconstruct the 3D imaging data at target timesteps ${\hat{\mathbf{X}}_{s_1},\hat{\mathbf{X}}_{s_2},...,\hat{\mathbf{X}}_{s_S}}$.

\subsubsection{Latent encoder: 3D-ConvGRU}
ConvGRU \cite{ballas2015delving} leverages convolutions within the GRU framework, enabling it to simultaneously process both spatial and temporal information in sequential data. Building on the concept of ConvGRU from \cite{ballas2015delving}, which employs 2D convolutions, this application utilizes 3D convolutions instead. This modified unit is referred to as 3D-ConvGRU and the corresponding update function is named 3D-ConvGRUCell. Given the sequence of quantized embeddings ${\mathbf{z_\mathbf{q}}_{t_0},\mathbf{z_\mathbf{q}}_{t_1},...,\mathbf{z_\mathbf{q}}_{t_T}}$, 3D-ConvGRU models the time series by making next-step prediction based on previous hidden state in an autoregressive way. The 3D-ConvGRU is run backwards as suggested by \cite{chen2018neural} and can be formulated as: $\mathbf{h}_{t_{i-1}}=\textrm{3D-ConvGRUCell}(\mathbf{h}_{t_i},\mathbf{z_\mathbf{q}}_{t_{i-1}})$, where $\mathbf{h}_{t_i}$ is the hidden state on $t_i$.

\subsubsection{Latent decoder}
The latent decoder comprises three components: a neural ODE, a 3D convolution layer, and a linear composition layer. This decoder architecture is adapted from \cite{park2021vid}.

The neural ODE defines a continuous hidden state $\mathbf{h}(t)$ which is the solution of an ODE initial-value problem (IVP) as follows \cite{chen2018neural,rubanova2019latent}. Here the initial status is $\mathbf{h}_{t_0}$ produced by the above 3D-ConvGRU.  
\begin{equation}
   \frac{d\mathbf{h}(t)}{dt} = f_{\theta}(\mathbf{h}(t), t), \; \; \; \; \; \; \; \mathbf{h}(t_0) = \mathbf{h}_{t_0}
\end{equation}

$f_\theta$ is a neural network parameterized by $\theta$ and $f_\theta$ defines the dynamics of $\mathbf{h}(t)$. By employing a numerical ODE solver, the hidden states ${s_1, s_2, ..., s_S}$ at any target timesteps can be obtained based on the initial status $\mathbf{h}_{t_0}$. This method excels by accommodating more complex dynamics within the latent state, as opposed to relying on restrictive assumptions like linearity \cite{sauty2022progression,kim2021longitudinal}. This enables more flexible disease progression modelling by using neural networks to directly parameterize the changes in the hidden state. Subsequently, a single 3D convolution layer takes two hidden states $\mathbf{h}_{s_i}$, $\mathbf{h}_{s_{i-1}}$ and outputs the difference map $\mathbf{D}_{s_i}$, approximating the difference $\Delta \mathbf{z_q}_{s_i}$ between the current $\mathbf{z_q}_{s_i}$ and the previous $\mathbf{z_q}_{s_{i-1}}$. 

The final output of the latent decoder $\hat{\mathbf{z_q}}_{s_i}$ is generated by combining previous generated output $\hat{\mathbf{z_q}}_{s_{i-1}}$ with the difference map. The loss for training the latent disease ODE comprises the sum of two components. The reconstruction loss, denoted as $\mathcal{L}_{\rm{recon}}$, is defined as $||\hat{\mathbf{z_\mathbf{q}}}-\mathbf{z_\mathbf{q}}||_2^2$, and the difference loss, denoted as $\mathcal{L}_{\rm{diff}}$, is calculated as $||\mathbf{D}_{s_i}-\Delta \mathbf{z_q}_{s_i}||_2^2$.

Feeding $\hat{\mathbf{z_\mathbf{q}}}_{s_i}$ to the trained decoder $G$ of 3D-VQ-GAN can reconstruct the corresponding 3D image $\hat{\mathbf{X}}_{s_i}$. The generative process of the decoder can be formulated as follows.

\begin{equation}
\begin{aligned}
    \mathbf{h}_{s_1},\mathbf{h}_{s_2},...,\mathbf{h}_{s_S} &= {\rm ODESolve}(\mathbf{h}_{t_0},f_{\theta},(s_1,...,s_S)),\\
&\mathbf{D}_{s_i}=\rm 3DConv(\mathbf{h}_{s_i},\mathbf{h}_{s_{i-1}}),\\
&\hat{\mathbf{z_q}}_{s_i} = \mathbf{D}_{s_i}+\hat{\mathbf{z_q}}_{s_{i-1}},\\
&\hat{\mathbf{X}}_{s_i}=G(\hat{\mathbf{z_q}}_{s_i})
\end{aligned}
\end{equation}

\subsection{Reconstruction performance of 3D-VQ-GAN}
\label{appendix:reconstruction}
% This experiment investigates the influence of two critical hyperparameters—codebook vocabulary size and compression rate—on the performance of the 3D-VQ-GAN model (Appendix~\ref{appendix:3dvqgan}). 
This experiment investigates the influence of two critical hyperparameters—codebook vocabulary size and compression rate—on the performance of the 3D-VQ-GAN model. The codebook vocabulary size specifies the number of discrete latent vectors in the codebook $\mathcal{Z}$, which serve as building blocks for representing input data in the latent space. A larger vocabulary size enables the model to capture finer details in 3D CT scans but increases computational demands. In contrast, the compression rate controls the degree of dimensionality reduction during encoding, reducing the complexity of the latent space representation. While a higher compression rate simplifies the model, it risks losing information and compromising reconstruction quality. To identify the optimal configuration, the model was trained on a training dataset and evaluated on internal and external test sets (Table~\ref{tab5_1}) using quantitative metrics (e.g., reconstruction error) and qualitative visual inspection. A vocabulary size of 256 and a compression rate of 4 were chosen as they offered the best balance between detail preservation, computational efficiency, and reconstruction performance. Figure~\ref{fig5_2} showcases examples of input 3D CT scans and the corresponding reconstructions generated by VQ-GAN models trained with different hyperparameter settings.

\subsection{Visualization of codebook}

\begin{figure}[H]
\includegraphics[width=0.8\textwidth]{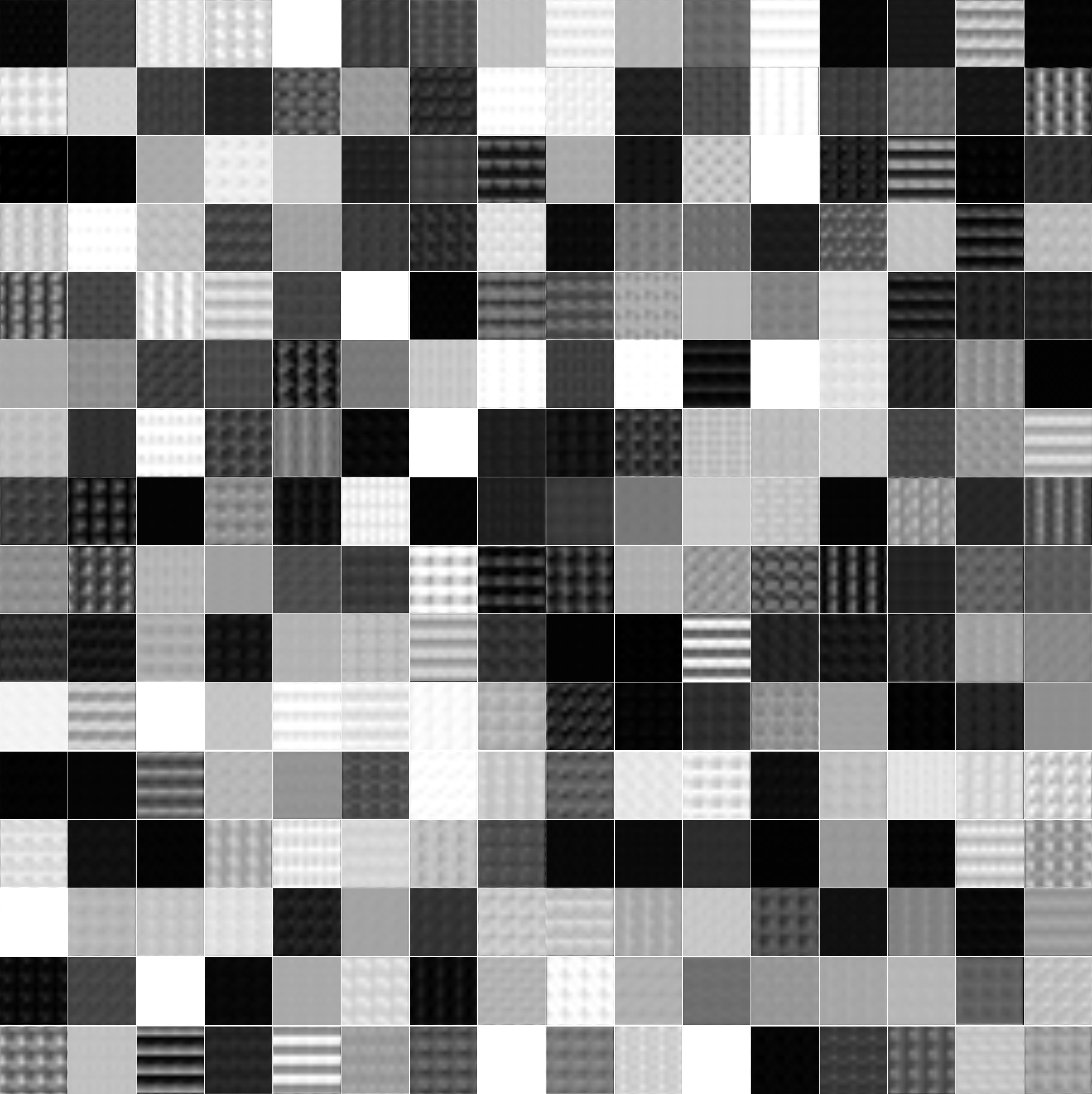}
\caption{A visualization of the codebook ($M=256$)} \label{fig5_4}
\end{figure}

Every entry in the codebook corresponds to a distinctive representation or code assigned to a specific region or pattern within the input space. Utilizing the techniques outlined in \cite{irie2023topological}, we visually represent each code in the learned codebook by creating a latent representation $\mathbf{z_q}$ using only that specific code. The resulting latent representation is then projected back into the 3D image space (see Figure \ref{fig5_4}). These visualizations exhibit varying grayscale intensities and textures, highlighting the diverse characteristics associated with different codes. This diversity within the codebook suggests that the model has effectively captured a broad array of features, enabling it to generate samples with varied and realistic qualities.

\newpage
\section{Related Work}
\subsection{Synthetic medical image generation}
Synthetic medical image generation proves particularly useful in various applications, classified into two types: unconditional and conditional, depending on whether constraints (e.g., images, a specific disease state, imaging modality, etc.) are applied respectively. The most common generative models used for image generation include Generative Adversarial Network (GAN) \cite{creswell2018generative}, VAE \cite{kingma2013auto}, and diffusion models \cite{ho2020denoising}. These models have demonstrated success in natural image generation and have shown their potential in the context of medical images. However, 3D imaging data (e.g., CT, MRI) is widely used in the medical field and generating realistic 3D images poses more significant challenges compared to 2D natural images due to the inherent complexity of three-dimensional space and the additional considerations required for realism. More specifically, unlike a 2D image with a single viewpoint, generating 3D images requires modelling the whole 3D structure which involves capturing depth information, spatial relationships, and fine details - essentially the "world behind the image". This requires not only higher computational resources but also more advanced techniques to model these detailed 3D structures with fidelity. The challenge is further amplified in the medical domain, where accurate representation of anatomical structures and simulation of physiological processes add significant layers of complexity.

Methods employing 3D-GANs have been proposed for the synthesis of 3D imaging data \cite{ferreira2022gan,singh2021medical}. However, training these models poses a considerable challenge due to increased computational and memory demands. In response to this issue, several memory-efficient 3D-GANs have been introduced \cite{sun2022hierarchical,uzunova2019multi}. GAN-based models are widely used in generating volumetric medical imaging data \cite{ferreira2024gan}. However, these models face additional challenges including mode collapse, non-convergence, and lack of interpretability. Mode collapse occurs when GANs fail to capture the full diversity of training data distribution and get stuck producing a limited set of outputs \cite{bau2019seeing}. Training GANs can also be difficult due to the need to balance and synchronize discriminator and generator. This often requires careful hyperparameter tuning and network architecture design to ensure convergence. Additionally, GANs typically lack interpretability, as it is challenging to understand what GANs have learned in the latent representation \cite{shen2020interfacegan}. In contrast, VAE \cite{kingma2013auto} has gained popularity for its explicit latent space representation and stable training process.

The Vector Quantized Variational Autoencoder (VQ-VAE) \cite{van2017neural}, a variant of VAE, was introduced to learn a discrete latent space, where continuous latent representations in the traditional VAE are quantized to discrete codes using a codebook. While the discrete latent space enhances efficiency and compactness, it also limits the model's ability to capture the full complexity of the input data, leading to blurry generated images. To address this limitation, VQ-VAE-2 \cite{razavi2019generating} utilized hierarchical multi-scale latent maps for large-scale image generation. VQ-GAN \cite{esser2021taming}, a variant of VQ-VAE, incorporated a discriminator and perceptual loss, combining the strengths of both VQ-VAE and GAN to generate high-resolution images. Ge et al. \cite{ge2022long} extended VQ-GAN for image modelling to 3D-VQ-GAN for video modelling. While pure GAN-based models dominate 3D medical image generation, VAE and VQ-VAE architectures are gaining traction. Existing applications focus primarily on brain and heart MRI scans \cite{liu20243d,tudosiu2020neuromorphologicaly}. Khader et al. \cite{khader2023transformers} demonstrated the potential of 3D-VQ-GAN for lung CT scans by combining it with transformers to generate realistic 3D CT scans based on a set of 2D radiographs. This highlights the capability of 3D-VQ-GAN for compressing volumetric lung CT scans.

Diffusion models \cite{ho2020denoising,croitoru2023diffusion} represent another emerging area in generative modelling. Diffusion models are a powerful class of probabilistic generative models that can learn complex distributions. These models initiate with a forward diffusion stage, where the input data is iteratively perturbed by adding noise, ultimately resulting in purely Gaussian noise. Subsequently, the models learn to reverse this diffusion process, aiming to reconstruct the original noise-free data from the noisy data samples \cite{kazerouni2023diffusion}. While diffusion models can generate diverse and high-quality images, their application in 3D imaging data synthesis remains underexplored due to their high computational cost and low sampling efficiency compared to VAE and GAN families. Khader et al. \cite{khader2022medical} employed a diffusion model in the lower-dimensional latent space of VQ-GAN rather than the image space to reduce computational costs and increase sampling efficiency.

\subsection{Temporal synthesis} 

Temporal synthesis can be viewed as a challenge that involves combining static image synthesis with temporal dynamics modelling. Recurrent Neural Networks (RNNs), Long Short-Term Memory (LSTM), and Gated Recurrent Unit (GRU) are frequently employed in temporal analysis. In addition, the recent success of transformer-based models in sequential data processing has sparked considerable interest due to their potential in modelling longitudinal data \cite{li2022hi}.

%While RNNs, LSTMs, and GRUs can handle irregularly sampled time series data, they are often more straightforward to use with regularly sampled data, which is not applicable in some applications. To better handle irregularly-sampled data, Rubanova et. al \cite{rubanova2019latent} proposed an ODE-RNN which generalizes the original RNNs to have continuous hidden dynamics governed by ODEs. 

Previous research on temporal synthesis often concentrated on video generation, with GAN-based models being the predominant methods inspired by the success of GANs in image generation. However, GAN-based approaches may face challenges in capturing long-term dependencies. Additionally, generating high-resolution frames or long video sequences presents difficulties due to prohibitively high memory and time costs during both training and inference \cite{ge2022long}. Some methods explore non-GAN-based generative models for video generation. Models presented in \cite{yan2021videogpt,ge2022long,le2021ccvs} employ VQ-VAE-based models and transformers for video generation, while \cite{ho2022imagen,voleti2022mcvd} utilize the diffusion model for video generation. Computational complexity, extended inference times, and temporal consistency remain open questions for these models. Other works, such as \cite{kanaa2021simple,park2021vid,xia2022modelling}, combine a typical encoder-decoder architecture with latent neural ODEs to capture temporal dynamics in the latent space for continuous-time video generation. 

Prior research in temporal synthesis for 3D imaging data has primarily focused on modelling two areas: normal brain ageing and disease progression in AD. This is often achieved by generating synthetic longitudinal brain MRIs. Normal ageing of the brain is characterized by a gradual loss of grey matter, particularly in the frontal, temporal, and parietal regions \cite{lorenzi2015disentangling}. In contrast, the brain morphology observed in AD patients reflects a combination of both normal ageing and pathological matter loss specific to the disease \cite{lorenzi2015disentangling}. These two processes can be modelled independently or jointly using temporal synthesis techniques \cite{sivera2019model}. Ravi et al. \cite{ravi2022degenerative} introduced the 4D-Degenerative Adversarial NeuroImage Net (4D-DANI-Net), a model crafted to generate high-resolution longitudinal MRI scans that replicate subject-specific neurodegeneration within the contexts of ageing and dementia. TR-GAN \cite{fan2022tr} was conceived to predict multi-session future MRIs based on prior observations, utilizing a single generator. Sauty et al. \cite{sauty2022progression} proposed a model that amalgamates a VAE with a latent linear mixed-effect model to estimate linear individual trajectories in latent space, enabling the sampling of patients’ trajectories at any given time point. While this model transforms observations at discrete time points into continuous disease progression trajectories, it relies on a strong assumption about linear trajectories in latent space. This linear assumption provides a simplified depiction of disease progression, but it falls short in capturing the inherent complexity observed in real-world disease dynamics. More flexible and adaptive approaches are needed to characterize disease progression trajectories effectively \cite{kim2021longitudinal}. To address this limitation, Martí-Juan et al. \cite{marti2023mc} employed a recurrent VAE where the latent space is parametrized with an RNN, defining more flexible disease evolution dynamics.

These temporal synthesis models for 3D imaging data offer substantial potential for clinical applications. These applications include: 1) Data Imputation: Longitudinal datasets are quite useful for the study of progressive diseases. However, longitudinal datasets often contain missing or incomplete data due to various reasons, such as missed appointments, dropout from the study, etc. \cite{fan2022tr} and \cite{fan2024tr} complement missing sessions for longitudinal MRI dataset expansion based on these models. 2) Assessing Treatment Efficacy: These models can create simulated longitudinal data that closely mimics the natural disease progression. This allows researchers to compare longitudinal imaging biomarkers between treated and untreated individuals at any point in time. By observing how the simulated disease course is altered by treatment, researchers can gain valuable insights into the treatment's effectiveness in slowing or even halting the disease process. This information can be crucial for designing future clinical trials and making informed treatment decisions. 3) Discovery of Temporal Biomarkers: Temporal synthesis models offer the ability to generate rich longitudinal imaging features. Analyzing the relationships between these features over time and how they connect to clinical outcomes can provide valuable insights. Researchers can leverage this approach to unlock the underlying mechanisms of disease progression and potentially discover novel temporal biomarkers.
% \newpage

% \begin{equation}\label{eq:example}
% \cos^2\theta + \sin^2\theta \equiv 1.
% \end{equation}

% \newpage

\end{document}